# On the Optimality of Beamforming for Multi-User MISO Interference Channels with Single-User Detection

Xiaohu Shang
Department of Electrical Engineering
Princeton University
Email: xshang@princeton.edu

Biao Chen
Department of EECS
Syracuse University
Email: bichen@syr.edu

H. Vincent Poor
Department of Electrical Engineering
Princeton University
Email: poor@princeton.edu

***Abstract*—For a multi-user interference channel with multi-antenna transmitters and single-antenna receivers, by restricting each receiver to a single-user detector, computing the largest achievable rate region amounts to solving a family of non-convex optimization problems. Recognizing the intrinsic connection between the signal power at the intended receiver and the interference power at the unintended receiver, the original family of non-convex optimization problems is converted into a new family of convex optimization problems. It is shown that, for such interference channels with each receiver implementing single-user detection, transmitter beamforming can achieve all boundary points of the achievable rate region.**

## I. Introduction

The interference channel (IC) models a multi-user communication system in which each transmitter communicates to its intended receiver while generating interference to all unintended receivers. Determination of the capacity region of an IC remains an open problem. To date, the best achievable rate region was established by Han and Kobayashi in [1], herein termed the HK region, which combines rate splitting at transmitters, joint decoding at receivers, and time sharing among codebooks. The HK region was later simplified in [2].

Recently, [3] showed that the HK region is within 1-bit of the capacity region of the Gaussian IC. The results in [4]–[6], whose genie-aided approach is largely motivated by [3], established the sum-rate capacity of the Gaussian IC in noisy interference: *treating interference as noise at both receivers is sum rate optimal*, i.e., each receiver should simply implement single-user detection (SUD). In addition, even if the noisy interference condition is not satisfied, practical constraints often limit the receivers to implementing SUD. For example, the receivers may know only the channels associated with their own intended links.

We assume in the present work that each receiver implements SUD, i.e., it treats interference as channel noise. In a preliminary work [7], we showed that beamforming is optimal for the entire SUD rate region for a two-user real multiple-input single-output (MISO) IC. This result was used in [8] to characterize the beamforming vectors that achieve the boundary rate points on the SUD rate region. Later, the result in [7] was also used in [9] to derive the noisy-interference sum-rate capacity of the symmetric real MISO ICs. We note that the proof in [7] is applicable only to a two-user real MISO IC.

There have been various studies concerning throughput optimization in a multi-user system under the assumption that *each receiver treats interference as channel noise*. However, even for the simple scalar Gaussian IC, computing the largest achievable rate region with SUD at each receiver is in general an open problem [10]. Exhaustive search over the transmitter powers is typically unavoidable due to the non-convexity of the problem. The difficulty is much more acute for the MISO IC case as one needs to exhaust all covariance matrices satisfying the power constraints, which renders the computation highly intractable. In this paper we propose an alternative way to derive optimal signaling for the SUD rate region for MISO ICs. Our approach is to convert a family of non-convex optimization problems for the original formulation to an equivalent family of convex optimization problems. What is more significant is that, given that each transmitter uses Gaussian input and each receiver implements SUD, *all boundary points of the rate region can be achieved by transmitter beamforming.*

The rest of the paper is organized as follows. In Section II, we prove that beamforming is optimal for the SUD rate region of a multi-user MISO IC. Numerical examples are provided in Section III. We conclude in Section IV.

Before proceeding, we introduce the following notation.

- Bold fonts $\boldsymbol{x}$ and $\mathbf{X}$ denote vectors and matrices respectively.
- $(\cdot)^T$ and $(\cdot)^\dagger$ denote respectively the transpose and the Hermitian (conjugate transpose) of a matrix or a vector.
- $\mathbf{I}$ is an identity matrix, $\mathbf{0}$ is an all zero matrix, and $\text{diag}(\cdots)$ is a diagonal matrix with its diagonal entries.
- $\mathbf{X} \succeq \mathbf{0}$ means that $\mathbf{X}$ is a symmetric positive semi-definite matrix.
- $\text{tr}(\mathbf{X})$ and $\text{rank}(\mathbf{X})$ denote the trace and the rank of matrix

[0]This research was supported in part by the National Science Foundation under Grant CNS-06-25637.

$\mathbf{X}$ respectively.
- $(\boldsymbol{x})_i$ denotes the $i$th entry of vector $\boldsymbol{x}$, and $\mathbf{X}_{m\times n}$ means that $\mathbf{X}$ is an $m \times n$ matrix.
- $\|\boldsymbol{x}\|$ is the norm of a vector $\boldsymbol{x}$, i.e., $\|\boldsymbol{x}\| = \sqrt{\boldsymbol{x}^\dagger \boldsymbol{x}}$.
- $E[\cdot]$ denotes the expectation.

## II. Multi-user MISO IC with single-user detector

We define the received signal for user $i$ of an $m$-user MISO IC as

$$Y_i = \sum_{j=1}^{m} \boldsymbol{h}_{ji}^T \boldsymbol{x}_j + N_i, \quad i = 1, \cdots, m, \tag{1}$$

where $\boldsymbol{x}_i$ is the transmitted signal vector of user $i$ with dimension $t_i$; $Y_i$ is the scalar received signal; the $N_i$ is unit variance Gaussian noise; and $\boldsymbol{h}_{ji}$ is the $t_i \times 1$ channel vector from the $j$th transmitter to the $i$th receiver. The power constraint at the transmitter is $\text{tr}(\mathbf{S}_i) \le P_i$, where $\mathbf{S}_i = E\left[\boldsymbol{x}_i \boldsymbol{x}_i^T\right]$. We assume that receiver $i$ knows only channel $\boldsymbol{h}_{ii}$, and decodes its own signal by treating the interference from all other users as noise. The boundary points of the achievable rate region for this channel is characterized by the following family of optimization problems:

$$\begin{aligned} \max \quad & \sum_{i=1}^{m} \mu_i R_i \\ \text{subject to} \quad & R_i = \frac{1}{2}\log\left(1 + \frac{\boldsymbol{h}_{ii}^T \mathbf{S}_i \boldsymbol{h}_{ii}}{1 + \sum_{j=1,j\neq i}^{m} \boldsymbol{h}_{ji}^T \mathbf{S}_j \boldsymbol{h}_{ji}}\right) \\ & \text{tr}(\mathbf{S}_i) \le P_i, \quad \mathbf{S}_i \succeq \mathbf{0}, \quad i = 1, \cdots m, \end{aligned} \tag{2}$$

where $0 \le \mu_i < \infty$.

Apparently problem (2) is a non-convex optimization problem. For each choice of $\boldsymbol{\mu} = [\mu_1, \cdots, \mu_m]$, all possible $[\mathbf{S}_1, \cdots, \mathbf{S}_m]$ must be exhausted. To obtain the entire SUD rate region, one has to go through this exhaustive search for all the $\boldsymbol{\mu}$ vectors.

Following the same problem reformulation procedure used in [7], to characterize the achievable rate region of $m$-user MISO IC, it is equivalent to solve the following family of convex optimization problems:

$$\begin{aligned} \max \quad & \boldsymbol{h}_{ii}^T \mathbf{S}_i \boldsymbol{h}_{ii} \\ \text{subject to} \quad & \boldsymbol{h}_{ij}^T \mathbf{S}_i \boldsymbol{h}_{ij} \le z_{ij}^2, \\ & \text{tr}(\mathbf{S}_i) \le P_i, \quad \mathbf{S}_i \succeq 0 \\ & i, j = 1, \ldots, m, \quad i \neq j, \end{aligned} \tag{3}$$

where $z_{ij}^2$ is a preselected constant denoting the interference power at the $j$th receiver caused by the $i$th transmitter. The problem reformulation is summarized in the following lemma.

*Lemma 1:* For any vector $\boldsymbol{\mu}$ with non-negative components, the optimal solution $\mathbf{S}_i^*$ for problem (2) is also an optimal solution for problem (3) with $z_{ij}^2 = z_{ij}^{*2} = \boldsymbol{h}_{ij}^T \mathbf{S}_i^* \boldsymbol{h}_{ij}$.

*Proof:* Problem (2) is equivalent to the following optimization problem for the same $\mu_i$:

$$\begin{aligned} \max \quad & \sum_{i=1}^{m} \frac{\mu_i}{2} \log\left(1 + \frac{\boldsymbol{h}_{ii}^T \mathbf{S}_i \boldsymbol{h}_{ii}}{1 + \sum_{j=1,j\neq i}^{m} z_{ji}^{*2}}\right) \\ \text{subject to} \quad & \boldsymbol{h}_{ij}^T \mathbf{S}_i \boldsymbol{h}_{ij} \le z_{ij}^{*2} \\ & \text{tr}(\mathbf{S}_i) \le P_i, \quad \mathbf{S}_i \succeq 0, \\ & i, j = 1, \cdots, m, \quad i \neq j. \end{aligned} \tag{4}$$

The equivalence is due to the following. First, the maximum of problem (2) is no smaller than that of problem (4), since problem (4) has extra constraint $\boldsymbol{h}_{ij}^T \mathbf{S}_i \boldsymbol{h}_{ij} \le z_{ij}^{*2}$. (This constraint is active, since the rates associated with $\mathbf{S}_{ii}^*$ are on the boundary of the SUD rate region.) On the other hand, the maximum of problem (2) is no greater than that of problem (4), since the $\mathbf{S}_i^*$'s are also feasible for problem (4). Therefore, problems (2) and (4) are equivalent. We now recognize that problem (4) is equivalent to problem (3) by setting $z_{ij} = z_{ij}^*$. ∎

We remark that the optimization problem (4) can not be solved directly as the constraint parameters $z_{ji}^*$ all depend on the unknown optimal covariances. That is, unless the optimal $\mathbf{S}_i^*$ of problem (2) is obtained, the equivalent optimization problem (4) cannot be parameterized. Although we do not give an explicit solution of problem (2) for a given $\boldsymbol{\mu}$ vector, Lemma 1 provides the following essential fact which is enough to obtain the entire SUD rate region defined by problem (2):

$$\bigcup_{\text{all } \mu_i, i=1,\cdots,m} \{\mathbf{S}_i^*(\mu_i)\} \subseteq \bigcup_{\text{all } z_{ji}, i,j=1,\cdots m, i\neq j} \{\mathbf{S}_i^*(z_{ji})\}, \tag{5}$$

where the left-hand side denotes the collection of the optimal solutions of problem (2) by exhausting $\boldsymbol{\mu}$, and the right-hand side denotes the collection of the optimal solutions of problem (3) by exhausting $z_{ij}$. Since the SUD rate region is determined by the left-hand side of (5), Lemma 1 successfully converts a family of non-convex optimization problems (2) into a family of convex optimization problems (4).

Based on Lemma 1, we obtain the following theorem.

*Theorem 1:* For an $m$-user MISO IC, the boundary points of the SUD rate region can be achieved by restricting each transmitter to implementing beamforming.

Theorem 1 can be readily extended to complex channels. Before proving Theorem 1, we first introduce the following lemma.

*Lemma 2:* Let $\boldsymbol{x}$ and $\boldsymbol{y}$ be two vectors with dimensions $t_1$ and $t_2$ respectively, and $\mathbf{K} \succeq \mathbf{0}$ be a $(t_1 + t_2) \times (t_1 + t_2)$ matrix with $\text{tr}(\mathbf{K}) \le P$. If

$$\mathbf{K} = \begin{bmatrix} \mathbf{K}_{11} & \mathbf{K}_{21}^T \\ \mathbf{K}_{21} & \mathbf{K}_{22} \end{bmatrix} \tag{6}$$

and $\mathbf{K}_{11} \succeq \mathbf{0}$ is a preselected $t_1 \times t_1$ matrix, then

$$\begin{bmatrix} \boldsymbol{x} \\ \boldsymbol{y} \end{bmatrix}^T \mathbf{K} \begin{bmatrix} \boldsymbol{x} \\ \boldsymbol{y} \end{bmatrix} \le \left(\sqrt{\boldsymbol{x}^T \mathbf{K}_{11} \boldsymbol{x}} + \|\boldsymbol{y}\| \sqrt{P - \text{tr}(\mathbf{K}_{11})}\right)^2, \tag{7}$$

and the equality can be achieved by choosing $\mathbf{K} = \mathbf{K}^*$:
1) When $\boldsymbol{x}^T \mathbf{K}_{11} \boldsymbol{x} \neq 0$ and $\|\boldsymbol{y}\| \neq 0$, we have

$$\mathbf{K}^* = \begin{bmatrix} \mathbf{K}_{11} & \frac{\sqrt{P - \text{tr}(\mathbf{K}_{11})}}{\|\boldsymbol{y}\| \sqrt{\boldsymbol{x}^T \mathbf{K}_{11} \boldsymbol{x}}} \mathbf{K}_{11} \boldsymbol{x} \boldsymbol{y}^T \\ \frac{\sqrt{P - \text{tr}(\mathbf{K}_{11})}}{\|\boldsymbol{y}\| \sqrt{\boldsymbol{x}^T \mathbf{K}_{11} \boldsymbol{x}}} \boldsymbol{y} \boldsymbol{x}^T \mathbf{K}_{11} & \frac{P - \text{tr}(\mathbf{K}_{11})}{\|\boldsymbol{y}\|^2} \boldsymbol{y} \boldsymbol{y}^T \end{bmatrix} \tag{8}$$

2) When $\boldsymbol{x}^T\mathbf{K}_{11}\boldsymbol{x} = 0$ and $\|\boldsymbol{y}\| \neq 0$, we have

$$\mathbf{K}^* = \begin{bmatrix} \mathbf{K}_{11} & \frac{\sqrt{P - \text{tr}(\mathbf{K}_{11})}}{\|\boldsymbol{y}\|}\mathbf{K}_{11}^{\frac{T}{2}}\mathbf{1}_0\boldsymbol{y}^T \\ \frac{\sqrt{P - \text{tr}(\mathbf{K}_{11})}}{\|\boldsymbol{y}\|}\boldsymbol{y}\mathbf{1}_0^T\mathbf{K}_{11}^{\frac{1}{2}} & \frac{P - \text{tr}(\mathbf{K}_{11})}{\|\boldsymbol{y}\|^2}\boldsymbol{y}\boldsymbol{y}^T \end{bmatrix}, \tag{9}$$

where $\mathbf{1}_0 = \begin{bmatrix} 1 \\ \mathbf{0}_{(t_1-1)\times 1} \end{bmatrix}$, $\mathbf{K}_{11}^{\frac{1}{2}} = \begin{bmatrix} \mathbf{\Lambda}^{\frac{1}{2}} & \mathbf{0} \\ \mathbf{0} & \mathbf{0} \end{bmatrix}\mathbf{Q}$,

with $\mathbf{K}_{11} = \mathbf{Q}^T \begin{bmatrix} \mathbf{\Lambda} & \mathbf{0} \\ \mathbf{0} & \mathbf{0} \end{bmatrix}\mathbf{Q}$

being the eigenvalue decomposition and $\mathbf{\Lambda}$ being a strictly positive diagonal matrix.

3) When $\|\boldsymbol{y}\| = 0$, we have

$$\mathbf{K}^* = \begin{bmatrix} \mathbf{K}_{11} & \mathbf{0} \\ \mathbf{0} & \mathbf{0} \end{bmatrix}. \tag{10}$$

For all three cases, we have

$$\text{rank}(\mathbf{K}^*) \leq \max\{\text{rank}(\mathbf{K}_{11}), 1\}. \tag{11}$$

The proof is omitted due to the space limitation.

Lemma 2 is useful for the following optimization problems:

$$\begin{aligned} \max \quad & \begin{bmatrix} \boldsymbol{x} \\ \boldsymbol{y} \end{bmatrix}^T \mathbf{K} \begin{bmatrix} \boldsymbol{x} \\ \boldsymbol{y} \end{bmatrix} \\ \text{subject to} \quad & h_i(\mathbf{K}_{11}) = 0, \quad i = 1, \cdots, n, \\ & g_j(\mathbf{K}_{11}) \leq 0, \quad j = 1, \cdots, m, \\ & \text{tr}(\mathbf{K}) \leq P, \quad \mathbf{K} \succeq \mathbf{0}, \end{aligned} \tag{12}$$

where $h_i(\cdot)$ and $g_j(\cdot)$ are fixed functions. By Lemma 2, we can convert the above problem into

$$\begin{aligned} \max \quad & \left(\sqrt{\boldsymbol{x}^T\mathbf{K}_{11}\boldsymbol{x}} + \|\boldsymbol{y}\|\sqrt{P - \text{tr}(\mathbf{K}_{11})}\right)^2 \\ \text{subject to} \quad & h_i(\mathbf{K}_{11}) = 0, \quad i = 1, \cdots, n, \\ & g_j(\mathbf{K}_{11}) \leq 0, \quad j = 1, \cdots, m, \\ & \text{tr}(\mathbf{K}_{11}) \leq P, \quad \mathbf{K} \succeq \mathbf{0}. \end{aligned} \tag{13}$$

Problems (12) and (13) have the same maximum. Once the optimal $\mathbf{K}_{11}$ for problem (13) is obtained, we can construct the optimal $\mathbf{K}$ for problem (12) by (8), (9) and (10). We note that the choices of (9) and (10) are not unique. One can choose $\mathbf{K}$ different from that of (8) and (9) and still achieve the same maximum of problem (12).

With Lemma 2, we prove Theorem 1 as follows.

*Proof:* By symmetry, it suffices to show that for the $m$th user, the optimal covariance matrix $\mathbf{S}_m^*$ for the following optimization problem satisfies rank$(\mathbf{S}_m^*) \leq 1$:

$$\begin{aligned} \max \quad & \boldsymbol{h}_{mm}^T\mathbf{S}_m\boldsymbol{h}_{mm} \\ \text{subject to} \quad & \boldsymbol{h}_{mj}^T\mathbf{S}_m\boldsymbol{h}_{mj} \leq z_{mj}^2, \quad j = 1, \cdots, m-1, \\ & \text{tr}(\mathbf{S}_m) \leq P_m, \quad \mathbf{S}_m \succeq \mathbf{0}, \end{aligned} \tag{14}$$

where all the $\boldsymbol{h}_{mj}$'s are $t_m \times 1$ vectors.

We first show that problem (14) can be written as

$$\begin{aligned} \max \quad & \left[\sqrt{\boldsymbol{h}^T\tilde{\mathbf{S}}_{11}\boldsymbol{h}} + \sqrt{\|\boldsymbol{h}_{mm}\|^2 - \|\boldsymbol{h}\|^2} \cdot \sqrt{P - \text{tr}(\tilde{\mathbf{S}}_{11})}\right]^2 \\ \text{subject to} \quad & \boldsymbol{h}_j^T\tilde{\mathbf{S}}_{11}\boldsymbol{h}_j \leq z_{mj}^2, \quad j = 1, \cdots, m-1 \\ & \text{tr}\left(\tilde{\mathbf{S}}_{11}\right) \leq P_m, \quad \tilde{\mathbf{S}}_{11} \succeq \mathbf{0}, \end{aligned} \tag{15}$$

where $\boldsymbol{h}$ and all the $\boldsymbol{h}_j$'s, $j = 1, \cdots, m-1$, are $\bar{m} \times 1$ vectors, $\tilde{\mathbf{S}}_{11}$ is an $\bar{m} \times \bar{m}$ matrix, and $\bar{m}$ is defined as

$$\bar{m} = \min\{t_m, m-1\}. \tag{16}$$

Obviously, when $\bar{m} = t_m \leq m-1$, problem (14) is exactly problem (15) by choosing $\boldsymbol{h} = \boldsymbol{h}_{mm}$, $\tilde{\mathbf{S}}_{11} = \mathbf{S}$ and $\boldsymbol{h}_j = \boldsymbol{h}_{mj}$. We need only to show the equivalence of problems (14) and (15) when $\bar{m} = m-1 < t_m$.

Let the singular value decomposition (SVD) of $\boldsymbol{h}_{m1}$ be

$$\boldsymbol{h}_{m1} = \mathbf{U}_1 \begin{bmatrix} \|\boldsymbol{h}_{m1}\| \\ \mathbf{0}_{(t_m-1)\times 1} \end{bmatrix},$$

and define

$$\mathbf{S}_m^{(1)} = \mathbf{U}_1^T\mathbf{S}_m\mathbf{U}_1 \tag{17}$$

$$\mathbf{h}_{mj}^{(1)} = \mathbf{U}_1^T\mathbf{h}_{mj}, \quad j = 1, \cdots, m. \tag{18}$$

Substituting (17) and (18) into (14), we obtain

$$\begin{aligned} \max \quad & \boldsymbol{h}_{mm}^{(1)T}\mathbf{S}_m^{(1)}\boldsymbol{h}_{mm}^{(1)} \\ \text{subject to} \quad & \boldsymbol{h}_{mj}^{(1)T}\mathbf{S}_m^{(1)}\boldsymbol{h}_{mj}^{(1)} \leq z_{mj}^2, \quad j = 2, \cdots, m-1, \\ & \begin{bmatrix} \|\boldsymbol{h}_{m1}\| \\ \mathbf{0}_{(t_m-1)\times 1} \end{bmatrix}^T \mathbf{S}_m^{(1)} \begin{bmatrix} \|\boldsymbol{h}_{m1}\| \\ \mathbf{0}_{(t_m-1)\times 1} \end{bmatrix} \leq z_{m1}^2, \\ & \text{tr}\left(\mathbf{S}_m^{(1)}\right) \leq P_m, \quad \mathbf{S}_m^{(1)} \succeq \mathbf{0}. \end{aligned} \tag{19}$$

Consider $\mathbf{h}_{m2}^{(1)}$ and let

$$\mathbf{h}_{m2}^{(1)} = \begin{bmatrix} \left(\mathbf{h}_{m2}^{(1)}\right)_1 \\ \left(\mathbf{h}_{m2}^{(1)}\right)_{2,\cdots,t_m} \end{bmatrix} = \begin{bmatrix} 1 & \mathbf{0} \\ \mathbf{0} & \mathbf{U}_2 \end{bmatrix} \begin{bmatrix} \left(\mathbf{h}_{m2}^{(1)}\right)_1 \\ \left\|\left(\mathbf{h}_{m2}^{(1)}\right)_{2,\cdots,t_m}\right\| \\ \mathbf{0}_{(t_m-2)\times 1} \end{bmatrix}, \tag{20}$$

where $\left(\mathbf{h}_{m2}^{(1)}\right)_{2,\cdots,t_m}$ is a vector consisting of the last $t_m - 1$ elements of $\mathbf{h}_{m2}^{(1)}$. The SVD of $\left(\mathbf{h}_{m2}^{(1)}\right)_{2,\cdots,t_m}$ is

$$\left(\mathbf{h}_{m2}^{(1)}\right)_{2,\cdots,t_m} = \mathbf{U}_2 \begin{bmatrix} \left\|\left(\mathbf{h}_{m2}^{(1)}\right)_{2,\cdots,t_m}\right\| \\ \mathbf{0}_{(t_m-2)\times 1} \end{bmatrix}, \tag{21}$$

where $\mathbf{U}_2^T\mathbf{U}_2 = \mathbf{I}_{(t_m-1)\times(t_m-1)}$. Therefore

$$\begin{bmatrix} 1 & \mathbf{0} \\ \mathbf{0} & \mathbf{U}_2 \end{bmatrix}^T \begin{bmatrix} 1 & \mathbf{0} \\ \mathbf{0} & \mathbf{U}_2 \end{bmatrix} = \mathbf{I}_{t_m \times t_m}. \tag{22}$$

Define

$$\mathbf{S}_m^{(2)} = \begin{bmatrix} 1 & \mathbf{0} \\ \mathbf{0} & \mathbf{U}_2 \end{bmatrix}^T \mathbf{S}_m^{(1)} \begin{bmatrix} 1 & \mathbf{0} \\ \mathbf{0} & \mathbf{U}_2 \end{bmatrix} \tag{23}$$

$$\mathbf{h}_{mj}^{(2)} = \begin{bmatrix} 1 & \mathbf{0} \\ \mathbf{0} & \mathbf{U}_2 \end{bmatrix}^T \mathbf{h}_{mj}^{(1)}, \quad j = 1, \cdots m. \tag{24}$$

On Substituting (20), (23) and (24) into (19), we have

$$\begin{aligned} \max \quad & \boldsymbol{h}_{mm}^{(2)T}\mathbf{S}_m^{(2)}\boldsymbol{h}_{mm}^{(2)} \\ \text{subject to} \quad & \boldsymbol{h}_{mj}^{(2)T}\mathbf{S}_m^{(2)}\boldsymbol{h}_{mj}^{(2)} \leq z_{mj}^2, \quad j = 3, \cdots, m-1, \end{aligned}$$

$$\begin{bmatrix}\|\boldsymbol{h}_{m1}\| \\ \mathbf{0}_{(t_m-1)\times 1}\end{bmatrix}^T \mathbf{S}_m^{(2)} \begin{bmatrix}\|\boldsymbol{h}_{m1}\| \\ \mathbf{0}_{(t_m-1)\times 1}\end{bmatrix} \le z_{m1}^2,$$
$$\begin{bmatrix}\left(\mathbf{h}_{m2}^{(1)}\right)_1 \\ \left\|\left(\mathbf{h}_{m2}^{(1)}\right)_{2,\cdots,t_m}\right\| \\ \mathbf{0}_{(t_m-2)\times 1}\end{bmatrix}^T \mathbf{S}_m^{(2)} \begin{bmatrix}\left(\mathbf{h}_{m2}^{(1)}\right)_1 \\ \left\|\left(\mathbf{h}_{m2}^{(1)}\right)_{2,\cdots,t_m}\right\| \\ \mathbf{0}_{(t_m-2)\times 1}\end{bmatrix}$$
$$\le z_{m2}^2,$$
$$\operatorname{tr}\left(\mathbf{S}_m^{(2)}\right) \le P_m, \quad \mathbf{S}_m^{(2)} \succeq \mathbf{0}. \tag{25}$$

We note that the above transformation does not change the form of the existing constraint (see the third lines of problems (19) and (25)). Now we continue the above procedure up to $\mathbf{h}_{m,m-1}$. In the $j$th transformation, we keep the first $j-1$ elements of $\mathbf{h}_{mj}^{(j-1)}$ and apply the SVD to the remaining $(t_m - j + 1)$ elements, and update the optimization problem. We formulate the $j$th iteration, $j = 2, \cdots, m-1$, as follows:

$$\boldsymbol{h}_{mj}^{(j-1)} = \begin{bmatrix}\left(\boldsymbol{h}_{mj}^{(j-1)}\right)_{1,\cdots,j-1} \\ \left(\boldsymbol{h}_{mj}^{(j-1)}\right)_{j,\cdots,t_m}\end{bmatrix}$$
$$= \begin{bmatrix}\mathbf{I}_{(j-1)\times(j-1)} & \mathbf{0} \\ \mathbf{0} & \mathbf{U}_j\end{bmatrix} \begin{bmatrix}\left(\boldsymbol{h}_{mj}^{(j-1)}\right)_{1,\cdots,j-1} \\ \left\|\left(\boldsymbol{h}_{mj}^{(j-1)}\right)_{j,\cdots,t_m}\right\| \\ \mathbf{0}_{(t_m-j)\times 1}\end{bmatrix},$$
$$\boldsymbol{h}_{mk}^{(j)} = \begin{bmatrix}\mathbf{I}_{(j-1)\times(j-1)} & \mathbf{0} \\ \mathbf{0} & \mathbf{U}_j\end{bmatrix}^T \boldsymbol{h}_{mk}^{(j-1)}, \quad k = 1, \cdots, m,$$
$$\mathbf{S}_m^{(j)} = \begin{bmatrix}\mathbf{I}_{(j-1)\times(j-1))} & \mathbf{0} \\ \mathbf{0} & \mathbf{U}_j\end{bmatrix}^T \mathbf{S}_m^{(j-1)} \begin{bmatrix}\mathbf{I}_{(j-1)\times(j-1))} & \mathbf{0} \\ \mathbf{0} & \mathbf{U}_j\end{bmatrix},$$

where $\left(\boldsymbol{h}_{mj}^{(j-1)}\right)_{j,\cdots,t_m}$ denotes the $j$th to the $t_m$th elements of $\boldsymbol{h}_{mj}^{(j-1)}$, and its SVD is

$$\left(\boldsymbol{h}_{mj}^{(j-1)}\right)_{j,\cdots,t_m} = \mathbf{U}_j \begin{bmatrix}\left\|\left(\boldsymbol{h}_{mj}^{(j-1)}\right)_{j,\cdots,t_m}\right\| \\ \mathbf{0}\end{bmatrix}, \tag{26}$$

where $\mathbf{U}_j^T \mathbf{U}_j = \mathbf{I}_{(t_m-j+1)\times(t_m-j+1)}$.

Finally, we convert problem (14) into the following form:

$$\begin{aligned} \max \quad & \begin{bmatrix}\boldsymbol{h} \\ \widehat{\boldsymbol{h}}\end{bmatrix}^T \tilde{\mathbf{S}} \begin{bmatrix}\boldsymbol{h} \\ \widehat{\boldsymbol{h}}\end{bmatrix} \\ \text{subject to} \quad & \begin{bmatrix}\boldsymbol{h}_j \\ \mathbf{0}\end{bmatrix}^T \tilde{\mathbf{S}} \begin{bmatrix}\boldsymbol{h}_j \\ \mathbf{0}\end{bmatrix} \le z_{mj}^2, \quad j = 1, \cdots m-1, \\ & \operatorname{tr}\left(\tilde{\mathbf{S}}\right) \le P_m, \quad \tilde{\mathbf{S}} \succeq \mathbf{0}, \end{aligned} \tag{27}$$

where $\boldsymbol{h}$ and $\boldsymbol{h}_j$ are $(m-1)\times 1$ vectors and $\widehat{\boldsymbol{h}}$ is a $(t_m - m + 1)\times 1$ vector. Furthermore, $\|\boldsymbol{h}_{mm}\|^2 = \|\boldsymbol{h}\|^2 + \left\|\widehat{\boldsymbol{h}}\right\|^2$. Let

$$\tilde{\mathbf{S}} = \begin{bmatrix}\tilde{\mathbf{S}}_{11} & \tilde{\mathbf{S}}_{21}^T \\ \tilde{\mathbf{S}}_{21} & \tilde{\mathbf{S}}_{22}\end{bmatrix}, \tag{28}$$

where $\tilde{\mathbf{S}}_{11}$ is an $(m-1)\times(m-1)$ matrix. The quadratic constraints in problem (27) are

$$\begin{bmatrix}\boldsymbol{h}_j \\ \mathbf{0}\end{bmatrix}^T \tilde{\mathbf{S}} \begin{bmatrix}\boldsymbol{h}_j \\ \mathbf{0}\end{bmatrix} = \boldsymbol{h}_j^T \tilde{\mathbf{S}}_{11} \boldsymbol{h}_j \le z_{mj}^2, \quad j = 1, \cdots m-1. \tag{29}$$

Therefore, the quadratic constraints in problem (27) are related only to $\tilde{\mathbf{S}}_{11}$. By Lemma 2, problem (27) is equivalent to problem (15).

We summarize that we have shown the equivalence of problems (14) and (15) with all the vectors in (15) being $\bar{m}\times 1$ and $\tilde{\mathbf{S}}_{11}$ being $\bar{m}\times\bar{m}$.

By Lemma 2, we can reconstruct $\tilde{\mathbf{S}}$ in a way such that $\operatorname{rank}\left(\tilde{\mathbf{S}}\right) \le \max\{\operatorname{rank}\left(\tilde{\mathbf{S}}_{11}\right), 1\}$. Let $\tilde{\mathbf{S}}_{11}^*$ be optimal for problem (27). To prove Theorem 1, it is equivalent to prove

$$\operatorname{rank}\left(\tilde{\mathbf{S}}_{11}^*\right) \le 1. \tag{30}$$

Furthermore, it suffices to prove that the rank of the optimal covariance matrix for the following optimization problem is no greater than 1:

$$\begin{aligned} \max \quad & \mathbf{h}^T \tilde{\mathbf{S}}_{11} \boldsymbol{h} \\ \text{subject to} \quad & \boldsymbol{h}_j^T \tilde{\mathbf{S}}_{11} \boldsymbol{h}_j \le z_{mj}^2, \quad j = 1, \cdots, m-1 \\ & \operatorname{tr}\left(\tilde{\mathbf{S}}_{11}\right) \le \bar{P}, \quad \tilde{\mathbf{S}}_{11} \succeq \mathbf{0}, \end{aligned} \tag{31}$$

where

$$\bar{P} = \operatorname{tr}\left(\tilde{\mathbf{S}}_{11}^*\right) \le P_m. \tag{32}$$

The equivalence is due to the fact that the optimal $\tilde{\mathbf{S}}_{11}$ for problem (15) is also optimal for problem (31) and vice versa.

Since $\tilde{\mathbf{S}}_{11} \succeq \mathbf{0}$, we can define

$$\tilde{\mathbf{S}}_{11} = \mathbf{B}^T \mathbf{B}, \tag{33}$$

where $\mathbf{B}$ is an $\bar{m}\times\bar{m}$ matrix. Then we can rewrite problem (31) as

$$\begin{aligned} \max \quad & \|\mathbf{B}\boldsymbol{h}\|^2 \\ \text{subject to} \quad & \|\mathbf{B}\boldsymbol{h}_j\|^2 \le z_{mj}^2, \quad j = 1, \cdots, m-1, \\ & \operatorname{tr}\left(\mathbf{B}^T\mathbf{B}\right) \le \bar{P}. \end{aligned} \tag{34}$$

The Lagrangian of problem (34) is

$$\begin{aligned} L = & -\|\mathbf{B}\boldsymbol{h}\|^2 + \sum_{j=1}^{m-1} \lambda_j \left(\|\mathbf{B}\boldsymbol{h}_j\|^2 - z_{mj}^2\right) \\ & + \lambda_m \left[\operatorname{tr}\left(\mathbf{B}^T\mathbf{B}\right) - \bar{P}\right]. \end{aligned} \tag{35}$$

Let $$\frac{\partial L}{\partial \mathbf{B}} = \mathbf{B}\left(\mathbf{C} + \lambda_m \mathbf{I}\right) = \mathbf{0}, \tag{36}$$

where $$\begin{aligned} \mathbf{C} & = -\boldsymbol{h}\boldsymbol{h}^T + \sum_{j=1}^{m-1} \lambda_j \boldsymbol{h}_j \boldsymbol{h}_j^T \\ & = \mathbf{H} * \operatorname{diag}\left[-1, \lambda_1, \cdots, \lambda_{m-1}\right] * \mathbf{H}^T \end{aligned} \tag{37}$$

where $\mathbf{H} = [\boldsymbol{h}, \boldsymbol{h}_1, \cdots, \boldsymbol{h}_{m-1}]$ is an $\bar{m}\times m$ matrix, and $\mathbf{C}$ is an $\bar{m}\times\bar{m}$ matrix.

We then introduce the following lemma which is an extension of Sylvester's Law of Inertia.

*Lemma 3:* [11, Theorem 7] Let $\mathbf{H}$ be an $m \times n$ matrix and $\mathbf{A}$ be an $n \times n$ Hermitian matrix. Denote $\pi(\cdot)$ and $\upsilon(\cdot)$ respectively as the numbers of positive and negative eigenvalues of a matrix. Then we have

$$\pi\left(\mathbf{HAH}^{\dagger}\right) \leq \pi\left(\mathbf{A}\right), \quad \upsilon\left(\mathbf{HAH}^{\dagger}\right) \leq \upsilon\left(\mathbf{A}\right).$$

By Lemma 3 and the Karush-Kuhn-Tucker (KKT) conditions that require $\lambda_i \geq 0$, $i = 1, \cdots, m$, we have

$$\pi(\mathbf{C}) \leq m - 1, \quad \upsilon(\mathbf{C}) \leq 1. \tag{38}$$

Since $\mathbf{C}$ is an $\bar{m} \times \bar{m}$ matrix, we can write the eigenvalue decomposition of $\mathbf{C}$ as

$$\mathbf{C} = \mathbf{Q}^T \text{diag}\left(\eta_1, \cdots, \eta_{\bar{m}}\right) \mathbf{Q} \tag{39}$$

where $\mathbf{Q}^T\mathbf{Q} = \mathbf{I}$, and $\eta_i$'s are the eigenvalues of $\mathbf{C}$ in ascending order. From (38), $\eta_1 \leq 0$ and $\eta_j \geq 0$, $j = 2, \cdots, \bar{m}$.

Since the optimal $\mathbf{B}^*$ satisfies tr $\left(\mathbf{B}^{*T}\mathbf{B}^*\right) = \bar{P}$, from the KKT conditions we have $\lambda_m > 0$. Thus, we have

$$\begin{aligned} &\text{rank}\left(\mathbf{C} + \lambda_m \mathbf{I}\right) \\ &= \text{rank}\left(\text{diag}\left[\lambda_m + \eta_1, \lambda_m + \eta_2, \cdots, \lambda_m + \eta_{\bar{m}}\right]\right) \\ &\geq \bar{m} - 1. \end{aligned} \tag{40}$$

Since $\mathbf{B}$ is an $\bar{m} \times \bar{m}$ matrix, from (36), we conclude that the optimal $\mathbf{B}^*$ for problem (34) satisfies

$$\text{rank}\left(\mathbf{B}^*\right) \leq 1. \tag{41}$$

Therefore

$$\text{rank}\left(\mathbf{S}_m^*\right) = \text{rank}\left(\tilde{\mathbf{S}}_{11}^*\right) = \text{rank}\left(\mathbf{B}^{*T}\mathbf{B}^*\right) \leq 1. \tag{42}$$

■

*Remark:* Theorem 1 proves the sufficiency of transmitter beamforming for achieveing the SUD rate region. However, it does not mean that the SUD rate region can only be achieved by beamforming. This depends on how we construct $\tilde{\mathbf{S}}$ after we obtain $\tilde{\mathbf{S}}_{11}$. In the proof we choose $\tilde{\mathbf{S}}$ as (8), (9) and (10) in Lemma 2. However, the choices of (9) and (10) are not unique. Another observation is that only (8) and (9) use full power. Equation (10) corresponds to the case that $\boldsymbol{h}_{mm}$ is linearly dependent of $\boldsymbol{h}_{mj}$, $j = 2, \cdots, m$, so that $\widehat{\boldsymbol{h}} = \mathbf{0}$ in (27). This agrees with the result for two-user scalar Gaussian IC in which the maximum SUD sum rate sometimes is achieved when one user is silent [12, Theorem 6].

## III. Numerical Examples

Using Theorem 1, we obtain in Fig. 1 the SUD rate region of a three-user MISO IC with the power constraint $P_1 = 1$, $P_2 = 1.5$ and $P_3 = 2$. The channels are

$$\mathbf{H}_1 = \begin{bmatrix} -2.1 & 0 & 0.5 \\ 0.1 & 0.2 & 0.1 \\ 1.5 & 0.9 & 0.3 \\ 0.1 & 0.2 & -1 \\ 0.2 & 0.8 & -0.9 \end{bmatrix}, \quad \mathbf{H}_2 = \begin{bmatrix} 0 & 2.7 & -0.5 \\ 0.4 & 0.4 & 0.2 \\ -0.9 & -1.3 & -0.6 \\ 0.8 & 0.4 & 0 \\ 0.1 & 0.5 & 0.4 \end{bmatrix}, \quad \mathbf{H}_3 = \begin{bmatrix} 1.2 & 0 & 1 \\ 0.8 & 0.9 & -1.7 \\ -2.6 & 0.8 & -1 \\ 0.3 & 1.3 & 0.7 \\ 0.8 & 1.2 & -1 \end{bmatrix},$$

where $\mathbf{H}_1 = [\boldsymbol{h}_{11}, \boldsymbol{h}_{12}, \boldsymbol{h}_{13}]$, $\mathbf{H}_2 = [\boldsymbol{h}_{21}, \boldsymbol{h}_{22}, \boldsymbol{h}_{23}]$ and $\mathbf{H}_3 = [\boldsymbol{h}_{31}, \boldsymbol{h}_{32}, \boldsymbol{h}_{33}]$. The solid curves are the rate regions for one user being inactive or at the maximum rate. That is, they are projections of the 3-D rate region onto a 2-D plane with one rate fixed at a constant value.

## IV. Conclusion

We have considered multi-user MISO ICs where each receiver is limited to single-user detection. By exploiting the relation between the signal power at the intended receiver and the interference power at the unintended receiver, we have converted the original family of non-convex optimization problems into an equivalent family of convex optimization problems. Transmitter beamforming is shown to be sufficient to achieve all boundary points of the SUD rate region.

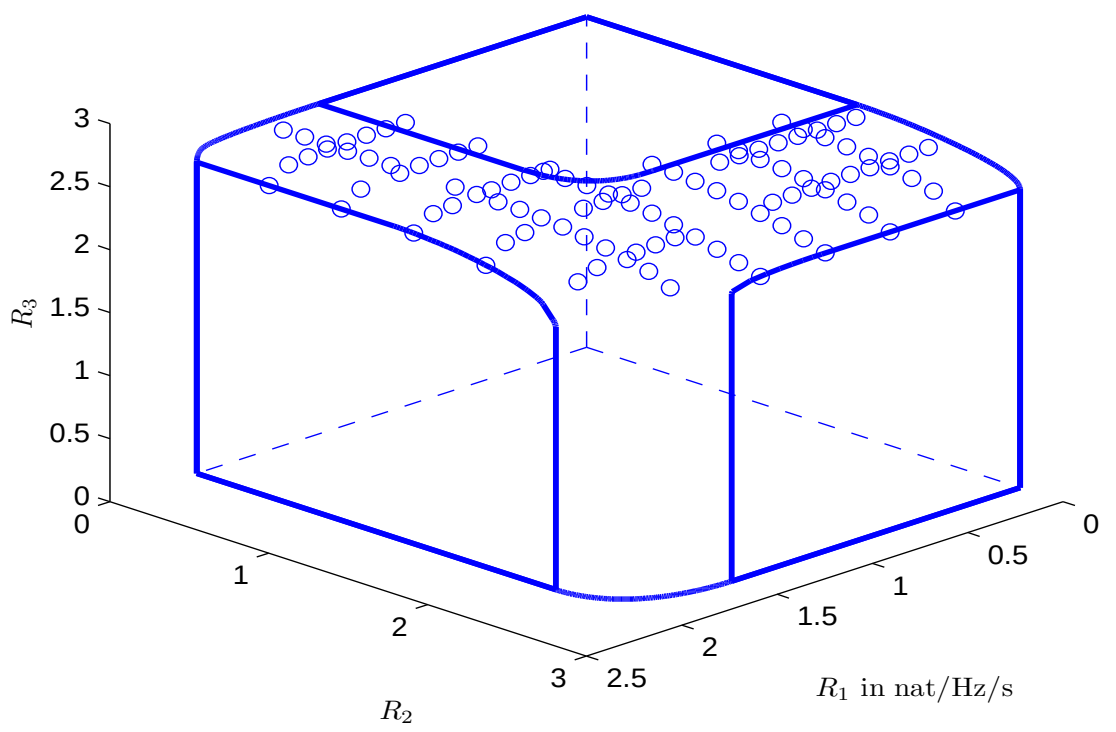

Fig. 1. The SUD rate regions of a three-user MISO IC.